\documentclass[a4paper,10pt,conference]{IEEEtran}
\usepackage[left=1.57cm,right=1.57cm,top=0.95cm,bottom=2.54cm]{geometry}

\usepackage[utf8]{inputenc}
\usepackage[T1]{fontenc}
\usepackage{textcomp}
\usepackage{stfloats}
\usepackage{microtype}
\usepackage{calc}
\usepackage[normalem]{ulem}
\usepackage{eurosym}
\usepackage{multirow} 

\usepackage[range-phrase=--,per-mode=symbol-or-fraction,binary-units=true,range-units=single,list-units=single,detect-all,forbid-literal-units]{siunitx}
\usepackage{silence}
\usepackage{csquotes}

\renewcommand{\footnoterule}{\kern -6pt}

\usepackage{amsmath}
\usepackage{amssymb}
\usepackage{amsfonts}
\usepackage{amsthm}

\usepackage{booktabs}
\usepackage{url}

\usepackage[inline]{enumitem}

\usepackage[caption=false,font=footnotesize]{subfig}
\usepackage{graphicx}
\usepackage{pgfplots}
\DeclareGraphicsExtensions{.pdf,.png,.jpg}

\usepackage{todonotes}

\usepackage{tikz}
\usetikzlibrary{arrows}
\usetikzlibrary{calc}
\usetikzlibrary{chains}
\usetikzlibrary{scopes}

\usepackage[american]{babel}
\usepackage{hyphenat}

\usepackage{algpseudocode}
\usepackage{algorithm}

\usepackage[capitalize,noabbrev]{cleveref}
\crefname{paragraph}{Paragraph}{Paragraphs}
\Crefname{paragraph}{Paragraph}{Paragraphs}

\usepackage[nodebug]{flushend}

\RequirePackage{xstring}
\RequirePackage{xparse}
\RequirePackage[]{acro}
\NewDocumentCommand\acrodef{mO{#1}mG{}}{\DeclareAcronym{#1}{short={#2}, long={#3}, #4}}
\acrodef{ITS}{intelligent transportation system}{short-plural-form={ITS}}
\acrodef{V2X}{vehicle-to-everything}

\usepackage{multirow}
\usepackage{comment}

\theoremstyle{plain}
\newtheorem{theorem}{Theorem}
\newtheorem{corollary}{Corollary}
\newtheorem{proposition}{Proposition}

\newtheorem{lemma}{Lemma}
\newtheorem{problem}{Problem}

\DeclareMathOperator{\E}{\mathbb{E}}

\def\bd{\begin{definition}}
\def\ed{\end{definition}}
\def\bt{\begin{theorem}}
\def\et{\end{theorem}}
\def\be{\begin{center}\begin{equation}}
\def\ee{\end{equation}\end{center}}
\def\bc{\begin{corollary}}
\def\ec{\end{corollary}}
\def\bl{\begin{lemma}}
\def\el{\end{lemma}}
\def\br{\begin{remark}}
\def\er{\end{remark}}
\begin{document}

\title{Assessing the Benefits of Ground Vehicles as Moving Urban Base Stations
} %
\author{
    \IEEEauthorblockN{Laura Finarelli}
    \IEEEauthorblockA{
        \textit{HES-SO Valais, Switzerland} \\
        \textit{TU Berlin, Germany} \\   
        laura.finarelli@hevs.ch
    }
    \and
    \IEEEauthorblockN{Falko Dressler}
    \IEEEauthorblockA{
        \textit{TU Berlin, Germany} \\        
        dressler@ccs-labs.org
    }
    \and
    \IEEEauthorblockN{Marco Ajmone Marsan}
    \IEEEauthorblockA{
        \textit{Institute IMDEA Networks} \\        
        marco.ajmone@imdea.org
    }
    \and
    \IEEEauthorblockN{Gianluca Rizzo}
    \IEEEauthorblockA{
        \textit{HES-SO Valais, Switzerland} \\
        \textit{Universita' di Torino, Italy} \\       
        gianluca.rizzo@hevs.ch
    }
}

\maketitle

\begin{abstract}
In the evolution towards 6G user-centric networking, the moving network (MN) paradigm can play an important role.
In a MN, some small cell base stations (BS) are installed on top of vehicles, and enable a more dynamic, flexible and sustainable, network operation. By "following" the users movements and adapting dynamically to their requests, the MN paradigm enables a more efficient utilization of network resources,  mitigating the need for dense small cell BS deployments at the cost of an increase in resource utilization due to wireless backhauling. This aspect is at least partly compensated by the shorter distance between users and BS, which allows for lower power and Line-of-Sight communications.
While the MN paradigm has been investigated for some time, to date, it is still unclear in which conditions the advantages of MN outweigh the additional resource costs.  
In this paper, we propose a stochastic geometry framework for the characterization of the potential benefits of the MN paradigm as part of an HetNet in urban settings. Our approach allows the estimation of user-perceived performance, accounting for wireless backhaul connectivity as well as base station resource scheduling. 
We formulate an optimization problem for determining the resource-optimal network configurations and BS scheduling which minimize the overall amount of deployed BSs in a QoS-aware manner, and the minimum vehicular flow between different urban districts required to support them, and we propose an efficient stochastic heuristic to solve it.
Our numerical assessment suggests that the MN paradigm, coupled with appropriate dynamic network management strategies, significantly reduces the amount of deployed network infrastructure while guaranteeing the target QoS perceived by users.  
\end{abstract}

\IEEEpeerreviewmaketitle

\section{Introduction}

The ever-increasing number of mobile subscribers, their insatiable appetite for data, and the surging demand for highly reliable, low-latency connectivity are pushing current cellular network paradigms to their limits. Traffic forecasts paint a daunting picture, with per-area data volume in future networks projected to surge by 1000 times compared to current levels \cite{Hurst2024UncrewedVI}, to support such high-bandwidth 6G applications as holographic communications or immersive virtual reality. This exponential growth, particularly pronounced in urban environments, necessitates innovative solutions to efficiently deliver high network capacity and a satisfying user experience.
Network densification, already a cornerstone of 5G technology \cite{densification}, addresses this challenge by deploying a denser network infrastructure of small cells. However, this approach comes at a significant cost. Densification puts a double whammy on network operators, driving up both capital expenditure (CAPEX) and operational expenditure (OPEX). Additionally, densification often relies on over-provisioning, deploying excess network resources to cope with unpredictable as well as periodic traffic fluctuations due to user movements (for example, from residential areas to business districts and back every working day). This approach, while ensuring network stability and high service availability, is inherently inefficient.
To overcome these limitations, unmanned vehicles are considered as key enablers of the 6G landscape, as they can address the limitations of traditional communication paradigms by enabling innovative use cases, enhancing network efficiency, improving system intelligence and resilience, and meeting the high demands of immersive applications \cite{Hurst2024UncrewedVI}.
Specifically, a promising avenue lies in the concept of \textit{moving networks} (MNs) \cite{Shan_survey_Mn_2021, finarelli2024mobile}. This paradigm is based on the integration of ground vehicles (GVs) as mobile base stations (MBSs), i.e., small cells mounted on vehicles or other mobile platforms, which can be strategically deployed to provide additional capacity exactly where and when needed, catering to localized traffic surges and thus potentially reducing over-provisioning needs.
The MN paradigm implies a much more extensive and flexible exploitation of the traditional approach based on truck-mounted BSs. While the latter were quasi-static, a MBS can follow users in their daily movement from a residential area to a business district, possibly also serving users stuck in a traffic jam while commuting together with the BS. By so doing, a mobile BS can take the place of several fixed small cell BSs, and take advantage of proximity to end users.

Previous studies have delved into optimizing MBS positioning and addressing mobility-related challenges. 
\cite{huang2019@resourceallocation2tierhetnet} proposes an optimization algorithm to ensure fairness among UEs while maximizing the throughput in two-tier networks with macro cells and cognitive micro cells. However, they do not account for the impact of backhauling.
A similar analysis with relay nodes, instead of mobile BS, was conducted as outlined in \cite{relay}. All these works however consider static scenarios, and thus do not account for  MBSs' ability to dynamically and naturally adapt BS density to both spatial and temporal variations in user density and traffic patterns, particularly in urban settings \cite{gupta2024densify}, thanks to the correlation between cell user densification and vehicular densification patterns. \cite{marsan2019towards} characterizes the correlation between patterns of mobile user densities and those of vehicles in a set of realistic urban scenarios. The authors also propose an approach for dynamic interference management in a network with MBSs. 
However, all these works do not allow quantifying the reduction in the amount of network resources required to serve users with a  target QoS which UGVs as MBS enable 
with respect to traditional,
static BS deployments (thus potentially reducing also the overall energy footprint of the network \cite{smallcellsEE}) and to unmanned aerial vehicles (UAVs) which consume more energy for the flying part.

This paper addresses this critical question by proposing a novel framework for deriving a first-order quantitative evaluation of the infrastructure savings achievable in MNs in Integrated Access and Backhaul (IAB) scenarios. We introduce an analytical approach to determine the optimal configuration in terms of static and mobile BS density aiming to minimize overall infrastructure requirements while guaranteeing the desired Quality of Service (QoS) for all users. Crucially, our approach accounts for the target QoS and resource utilization of wireless backhauling links. 
Our main contributions are:
\begin{itemize}[leftmargin=8pt]
    \item We propose an analytical approach for the analysis of moving networks, which incorporates the effects of IAB of moving BSs;
    \item We elaborate an optimization framework for the derivation of the CAPEX optimal configurations of the network, while ensuring a target QoS for users and backhaul;
    \item We characterize the optimal configurations resulting from our optimization framework in a simplified urban baseline scenario, allowing a first evaluation of those conditions in which the moving network paradigm enables substantial savings in terms of deployed infrastructure.
\end{itemize}
\section{System model and Assumptions}
\label{sec:System model}

We consider a finite area of the plane modeling an urban scenario, populated by BSs belonging to two tiers. Namely, we assume a fraction of the BSs in the area, denoted as moving BS (MBS), are installed on vehicles, such as cars or drones, and thus able to move between regions, while the remaining BSs, denoted as static BS (SBS), are at fixed locations, both distributed according to homogeneous planar Poisson Point Processes (PPP). At any time instant, we assume MBS distributed uniformly at random in the given area. The mobility of MBSs is modeled such that their spatial distribution varies dynamically across different time slots. An analogous modeling approach applies to user dynamics.
We consider user equipments (UEs) are broadband (BB) terminals, distributed in space according to a PPP. Note though that our analysis extends straightforwardly to scenarios with heterogeneous users, e.g. including IoT devices. 
The two PPPs that describe the distribution in space of MBSs and UEs are assumed to be positively correlated (as observed in \cite{adaptivedensification}) so that the system can deploy MBS since UEs are present in the scenario.
We define the \textit{observation window} as a finite time interval divided into $J$ equal-sized sections, where $j \in 1,..., J$ denotes the label of the $j$-th section.
We assume the considered area to be partitioned into $Z$ different regions with specific traffic profiles. In every time interval $j$, $\lambda^{z}_{u,j}$ denotes the intensity of the PPP of UEs, in users per $m^2$.
Such intensity might vary between intervals to model day/night patterns of activity or commuting, among others.\\
 We assume in-band wireless backhauling for MBSs, by which SBSs act as wireless access points for MBSs.
We adopt a channel model that considers only distance-dependent path loss, though our results can be easily extended to incorporate fading and shadowing. We assume a \textit{random frequency reuse} scheme, with reuse factor $k$. 
We consider only the downlink of the cellular access, and we assume the generic UE (respectively, MBS) located at $x$ is served by the BS (respectively, SBS) that provides the largest SINR at $x$. Since we are ignoring fading effects, we can reasonably assume that users are served by the BS providing the highest received power. \cite{Baccelli_stochasticgeometry}. 
We assume that static base stations correspond to macro cells, which transmit at a power level $P_s$, while moving base stations are modeled as small cells transmitting at a power level $P_m \leq P_s$.
Leveraging Shannon's capacity law, the capacity of a user located at a distance $r$ from the BS is $C(r,P,I) = (B/k)\log_{2}(1 + P r^{-\alpha}(N_0+I(r,k))^{-1})$, 
where $B$ is the channel bandwidth, $\alpha$ is the attenuation coefficient, $I$ is the total received interfering power, $N_0$ is the power spectral density of the additive white Gaussian noise, and $P$ is the transmission power of the serving base station tier.\\
We assume that a weighted processor-sharing (WPS) mechanism is used to divide BS time among all the connected users. By doing so, a notion of fairness is imposed among users from a same class and associated to a same BS, since they are all  served for an identical fraction of time.
 We model wireless inband backhauling by assuming that SBSs serve two classes of users (BB users and MBS), each with its own WPS weight.\\ 
To model the QoS perceived by a user, we use the \textit{per-bit delay}, defined as the inverse of the short-term user throughput \cite{delay1}. We assume that the mean utilization of BS (i.e., the fraction of time during which a BS, mobile or static, is active) and the WPS weights are tuned in such a way as to achieve the target QoS for all classes of users.\\
We assume that at each SBS the WPS weights for MBSs and BB users are equal to one, and $\phi_j^z\geq 0$, respectively, with $\phi_j^z$ taking the same value for all SBSs.
The utilization of a SBS serving $N_{M}$ MBSs and $N_{U}$ UEs is thus $U_s=({N_{M}+\phi_j^z N_{U}})/({N_{M}+\phi_j^z N_{U}+\beta_s})$, where $\beta_s$ is the fraction of BS time during which the SBS is not serving any user (and thus potentially saving energy). For MBS instead, we have $U_m={N_{U}}/({N_{U}+\beta_m})$\label{util_MBS}. For BB users, the key performance parameter is the Palm expectation of the per-bit delay experienced by a typical user who is just beginning service \cite{rizzo2015}. For backhauling links instead, the main performance indicator is the \textit{violation probability}, i.e. the probability that the per-bit delay perceived by the backhauling connection is insufficient to carry the aggregate traffic demand.\\
Thus, both MBSs and SBSs tune their utilization and the WPS weight to have at the same time the Palm expectation of the per-bit delay perceived by the typical BB user coinciding with a target value $\tau_0$, and the violation probability for BH traffic below a target value $\delta$.

\section{An analytical model for user-perceived performance}
\label{sec:analytical_model}

To derive analytical relationships between the main system parameters and the network's key performance indicators, we focus on the \textit{ideal per-bit delay} perceived by a user, which is the per-bit delay perceived when the serving BS has utilization equal to $1$. Indeed, given our assumptions, the actual per-bit delay is given by the product of the ideal per-bit delay and the BS utilization.\\ 
Let $S(x)$ denote the location of the BS that is closest to the user located at $x$. For the typical user at the origin served by the closest SBS in $S(0)$, the ideal per-bit delay perceived is
\begin{equation}\label{eq:user_sbs}
   \tau_s(S(0)) = \frac{N_M(S(0))+\phi_j^z N_{U}(S(0)) }{\phi_j^z C(D(0), P_s,I)} 
\end{equation}
where $N_M$ and $N_{U}$ are the number of MBSs and UEs served by the SBS in $S(0)$, respectively, and $D(0)$ is the distance between the user at the origin and its serving BS.
If the user at the origin is an MBS (i.e. if it is a backhauling connection), the ideal per-bit delay is $\tau_M(S(0)) = \tau_s(S(0))\phi_j^z$. Similarly, if a MBS serves the BB user, the ideal per-bit delay the user would perceive is 
\begin{equation}\label{eq:tmd}
   \tau_m(S(0)) = \frac{N_{U}(S(0)) }{C(D(0),P_m,I)} 
\end{equation}
 where $N_{U}$ is the number of UEs served by the MBS in $S(0)$, and $D(0)$ is the distance between the given MBS and its serving SBS.\\
In region $z$ and time interval $j$, let $\bar{\tau}_{m}^{j,z}$ and  $\bar{\tau}_{s}^{j,z}$ denote the Palm expectation of the ideal per-bit delays perceived by BB users joining the system, and $\lambda^{z}_{m,j}$ (resp. $\lambda^{z}_{s}$) the mean density of MBSs (resp. of SBSs, which remains constant over time). 
With the following result, we model the probability distribution function of the distance between the users and their serving base station in our HetNet with two BS tiers. 

\begin{lemma}\label{lemma:pdf_distance}
Given a heterogeneous network with two populations of BSs, denoted as $m$ and $s$, both distributed as a PPP with intensity $\lambda_i$ and transmit power $P_i$, $i\in m,s$. Then the pdf of the distance of the ideal user arriving in the system from its serving base station is 
\begin{equation}\label{pdf_distance}
f_{m,s}(r)=2\pi r  (\lambda_s + \lambda_m\rho_{ms}^2) e^{-\pi r^2 (\lambda_s+\lambda_m \rho_{ms}^2)}
\end{equation}
with $\rho_{ms}=\sqrt[\alpha]{\frac{P_m}{P_s}}$.
\end{lemma}
\noindent For the proof, please refer to 
\cref{app:proof_pdf_distance} of the extended version \cite{IEEE_medcom_2025}.\\
The following result connects the mean total interference perceived by a user to the mean ideal per-bit delay perceived by users in the system.
\begin{lemma}\label{lemma:interference}
The mean total interference power perceived by a user at a distance $r$ from its serving BS in $z$ and interval $j$ is
    \begin{equation*}
    \bar{I} (r,\bar{\tau}_{m}^{j,z},\bar{\tau}_{s}^{j,z}) = \frac{2\pi r^{2-\alpha} }{k(\alpha-2)\tau_0} (P_m \bar{\tau}_{m}^{j,z}\lambda_{m,j}^z +P_s  \bar{\tau}_{s}^{j,z}\lambda_{s,j}^z ) 
\end{equation*} 
\end{lemma}
\noindent For the proof and the derivation of the average BS utilization, please refer to 
\cref{app:proof_lemma_interference}  of the extended version \cite{IEEE_medcom_2025}.\\
The following theorem gives the expressions for the Palm expectation of the ideal per-bit delays perceived by BB users.
\begin{theorem}\label{th:mean_tau}
    In a HetNet with two tiers of BS, moving and static, the Palm expectation of the ideal per-bit delay perceived by BB users joining the system in region $z$ and time interval $j$, and served by a MBS (respectively, a SBS) is approximated as the unique solution of the following fixed point problem:
 \begin{equation}\label{eq:tau_moving}
\begin{aligned}
        \bar{\tau}_{m}^{j,z}  &= \int_0^\infty \frac{\lambda_{u,j}^z h_m(r)}{1-e^{-\lambda_{u,j}^zh_m(r)}} H(P_m,r) dr \\
        \bar{\tau}_{s}^{j,z} &= \int_0^\infty \biggl(\frac{\phi_j^z\lambda_{u,j}^z h_s(r)}{1-e^{-\lambda_{u,j}^z h_s(r)}}+\lambda_{m,j}^z h_{BH}(r)\biggr) \frac{H(P_s,r)}{\phi_j^z}  dr 
        \end{aligned}
    \end{equation}
  with  
  \[
H(P, r) = 
\begin{aligned}
    &\frac{e^{-(\rho_{ms}^2\lambda_{m,j}^z + \lambda_{s}^z)\pi r^2}
    (\rho_{ms}^2\lambda_{m,j}^z + \lambda_{s}^z) 2\pi r}
    {C(r, P, I(r, \bar{\tau}_{m}^{j,z}, \bar{\tau}_{s}^{j,z}))}
\end{aligned}
\]
 \begin{equation}\label{eq:h_r}
 \begin{aligned}
h_m(r)&=
\int_0^\infty \int_0^{2\pi} e^{-\lambda_{s}^z A(r,x/\rho_{ms},\theta) - \lambda_{m,j}^z A(r, x,\theta)}x d\theta dx\\
 h_s(r)&= 
 \int_0^\infty \int_0^{2\pi} e^{-\lambda_{s}^z A(r,x,\theta) - \lambda_{m,j}^z A(r, x\rho_{ms},\theta)}x d\theta dx\\
h_{BH}(r)&=
\int_0^\infty \int_0^{2\pi} e^{- \lambda_{s}^z A(r,x,\theta)} x d\theta dx
\end{aligned}
\end{equation}
and $A(r, x, \theta)= \pi x^2 - 
[r^2 \arccos\left(\frac{r +  x \sin(\theta)}{d(r, x, \theta)}\right)+$

$+x^2\arccos \left(\frac{x + r \sin(\theta)}{d(r, x, \theta)}\right)-\frac{1}{2} \sqrt{r^2-(d(r, x, \theta)-x)^2 }\cdot$
$\cdot\sqrt{(d(r, x, \theta) +x)^2-r^2}]$, where 
$d(r,x,\theta)$ is the euclidean distance between $(x,\theta)$ and $(0,-r)$.
\end{theorem}
\noindent For the proof, please refer to 
\cref{app:proof_th_mean_delay}  of the extended version \cite{IEEE_medcom_2025}. Note that these results assume that there is at least one user in every cell. This is reasonable in practical scenarios, as when user density is lower than this (off-peak, e.g. at night), operators typically put some BSs to sleep to save energy, or in the case of MBS, they move them where needed.


The following result gives the expressions for the $CDF$ of the per-bit delay in backhauling. 
\begin{theorem}
\label{th:cdf}
In a given time interval $j$ and region $z$, the $CDF$ of the ideal per-bit delay which an SBS delivers to an MBS, denoted as $\tau_{M}$, is given by $CDF_r \Bigl( g^{-1}(\tau) \Bigr)$, with
$$CDF_r(y)= \int_0^y e^{-\lambda_{s}^z \pi x^2}\lambda_{s}^z 2 \pi x dx$$
and
$g(r) = \frac{\lambda_{m,j}^z h_{BH}(r)+\phi_j^z \lambda_{u,j}^z h_s(r)}{C(r,P_s,I(r, \bar{\tau}_{m}^{j,z},\bar{\tau}_{s}^{j,z}))}$.
\end{theorem}
\noindent For the proof, please refer to 
\cref{app:proof_th_cdf}  of the extended version \cite{IEEE_medcom_2025}.
To model the violation probability of a backhauling link, we focus on the traffic demand at a MBS, in terms of the per-bit delay of the aggregate downlink traffic at the given MBS. 
Then we express the violation probability as the probability that the per-bit delay perceived in downlink by MBSs is superior to the aggregate traffic demand in downlink of an MBS.  
\begin{theorem}\label{thm_violprob}
    The BH violation probability in region $z$ and interval $j$ is given by:
           \begin{equation*}
        V_{j}^z=
        \int_0^\infty f_{d}\left(\frac{\bar{\tau}_{s}^{j,z}x}{\tau_0} \right)\left(1-CDF_{\tau_M}(x-t)\right)dx
    \end{equation*}
    with
\begin{equation*}
    f_{d}(\tau) = \left(\frac{\tau_0}{\lambda_{u,j}^z}\right)^{\frac{7}{2}}\frac{343}{15}\sqrt{\frac{7}{2\pi}}\tau^{-\frac{9}{2}}e^{-\frac{7\tau_0}{2\lambda_{u,j}^z\tau}}
\end{equation*}    
     %
\end{theorem}
\noindent For the proof, please refer to 
\cref{app:proof_th_1} of the extended version \cite{IEEE_medcom_2025}.

\section{Problem formulation}
\label{sec:Problem formulation}

In this section, we present the formulation of the optimization problem, which provides, for a given mean user density, the density of BSs, both mobile and static, and the round-robin coefficients which minimize the overall network deployment costs while accounting for the target QoS for both fronthaul and wireless backhaul traffic.\\
To model the benefit of having BSs that move following the users' densification patterns, we assume that the total number of MBS in the area is constant over time. This approximates well actual mobility patterns in a city, when the considered urban area is large enough to cover the bulk of the daily commuting patterns.
Thus we have $\sum_{z=1}^Z\lambda^{z}_{m,j}E_z=M$ $\forall\;j$, where $E_z$ denotes the area of the $z-$th region. 

\begin{problem}{\textbf{Minimization of BS deployment costs}}\label{Prob:Opt_1}
\[
\underset{\lambda_{m,j}^z,\lambda_{s}^z, \phi_j^z}{\text{minimize}}\   \mu M+\sum_{z=1}^Z \lambda^{z}_{s} E_z 
 \]
Subject to, $\forall j,\;z$:
	\begin{align}
 \sum_{z=1}^Z\lambda^{z}_{m,j}E_z&=M\\
\bar{\tau}_{m,j}^z, \bar{\tau}_{s,j}^z & \leq \tau_0\label{constraint_um}\\
V_{j}^z & \leq \delta\label{constraint_violation}\\
\phi_j^z & \geq 0 
	\end{align}
\end{problem}
Parameter $\mu$ captures the relative difference in unitary cost (purchase and deployment) between SBSs and MBSs.
Constraints \ref{constraint_um} impose that, both in MBSs and in SBSs, the target QoS for BB users is achieved. Indeed, a Palm expectation of the ideal per-bit delay perceived by users larger than the target value $\tau_0$ implies that even with utilization equal to one, BSs do not possess enough resources to guarantee the target QoS for BB users. The value of $\delta$ in \ref{constraint_violation} must be set to a low enough value for violation events to have a negligible effect on the performance perceived by users.\\

\subsection{GA-Based Metaheuristic}
\label{mh}
Problem 1 thus aims at minimizing overall network deployment costs while achieving the target QoS for both BB users and BH links. Problem 1 has nonlinear and nonconvex constraints, thus it cannot be solved efficiently. To address this, we propose a metaheuristic 
whose fundamental idea is to maximize the amount of reused base stations, to compensate for the extra resource costs of MBS due to backhauling. Our heuristic goes through three steps:
\begin{itemize}[leftmargin=*]
     \item Step 1: In every region, a lower bound to SBS density is the minimum of the overall BS density required to serve users in the whole observation window, as those BS do not need to move between regions. These lower bounds (denoted as $\lambda_{\text{OnlyStatic}}^z$) are derived by solving Problem \ref{Prob:Opt_1} with $\lambda_{m,j}^z = 0$ $\forall j,z$ via the hippopotamus algorithms (HA) \cite{ho}, which has proven more efficient than the classical genetic algorithm (GA). The inequality constraints of Problem 1 are incorporated through an additive penalty objective function.
    \item Step 2: Problem 1 is solved with HA independently for every region $z$, and assuming $\lambda_{s}^z\geq\lambda_{\text{OnlyStatic}}^z$.
  Let $\lambda_{s}^{z*}$, $\lambda_{m,j}^{z*}$ and $\phi_j^{z*}$ denote the output of this step. 
    \item Step 3: The values of $\lambda_{m,j}^{z*}$ found in step 2 represent an upper bound on the optimal densities of MBS in every region and time interval. This is because when the original Problem 1 is considered, possibly not all MBS in excess in a given region and time interval may be reused in other regions in that time interval, due to e.g. low traffic demand in those regions at the time interval in which that excess is present. From the solutions of step 2, for every $z$ let $\Delta_{m}^z=min_j \lambda_{m,j}^{z*}$ denote the fraction of MBS which, in the solutions of step 2, is present in every time interval $z$ (and which therefore does not need to move).
    Thus in step 3 Problem 1 is solved with the additional constraints $\forall z$, $\lambda_{s}^{z*}\leq \lambda_{s}^z\leq\lambda_{s}^{z*}+\Delta_{m}^z$, and $0\leq\lambda_{m,j}^{z}\leq\lambda_{m,j}^{z*}$. Indeed the optimal SBS density is lower than $\lambda_{s}^{z*}+\Delta_{m}^z$, as when the excess MBS are swapped with an equivalent amount of SBS there are fewer MBSs to serve, and the backhauling traffic load is shared among a larger set of SBS. These two constraints greatly reduce the solution space with respect to the original problem formulation, drastically improving the computational efficiency and accuracy of the HA algorithm.   
\end{itemize}
\section{Numerical assessment}
\label{numerical results}
%
%
\begin{table}
  \centering
  \caption{Analytical vs simulation results for $\tau_0 = 10^{-3}$ s, $10$ W transmit power, and $50\%$ moving BSs.}
\scalebox{0.8}{
  \begin{tabular}{llccc}
    \toprule
    User density $[m^{-2}]$  & & $10^{-3}$ & $10^{-2}$ & $10^{-1}$\\
    \midrule
    Total BS density $[m^{-2}]$ & &  $1.79e{-4}$ & $6.78e{-4}$ & $1.2e{-3}$ \\
     \midrule
     \multirow{ 3}{*}{\parbox[c]{2.5cm}{\raggedright \textbf{Mean per-bit delay  (SBS)} $[\mu s]$}} & Analysis & $334$ & $150$ & $274$ \\
    &  Simulation & $348$ & $147$ & $289$ \\
    &  $95\%$ Conf. Int. & $[238, 458]$ & $[133, 160]$ & $[273, 304]$ \\
   \midrule
     \multirow{ 3}{*}{\parbox[c]{2.5cm}{\raggedright \textbf{Mean per-bit delay  (MBS)} $[\mu s]$}} & Analysis & $437$ & $403$ & $350$ \\
    & Simulation & $428$ & $401$ & $352$ \\
    & $95\%$ Conf. Int. & $[377, 476]$ & $[391, 412]$ & $[345, 358]$ \\
    \midrule
    \textbf{Violation Probability} & Downlink & 1.55\% & 0\% & 0.56\% \\
   \bottomrule
  \end{tabular}}
  \label{tab_sim}
  \vspace{-1em}
\end{table}
%
In this section, we validate numerically our analytical results and we investigate the potential resource savings enabled by the moving network paradigm as a function of the main system parameters.
%
We assume all BSs work at a frequency of $1.5$ GHz and use a bandwidth of $10$ MHz.
We consider a target QoS for BB users of $10^{-5}$ s (corresponding to a mean throughput of $100$ kbps)
and of $10^{-6}$ s (1 Mbps).
We assume a path-loss coefficient of $3$, typical of urban scenarios, a $3$ W transmit power and a reuse factor of $3$, as typical of small cell settings \cite{gupta2024densify}. 
We consider a density of connected users ranging from $10^3$ to $10^5$ users per $km^2$, as close to typical daily variations in residential and business districts during working hours in many large cities. 
\\
The multiplicative factors in the penalty functions used to solve Problem 1, derived empirically, have been $150$ for Constraint \ref{constraint_um}.1, $100$ for Constraint \ref{constraint_um}.2, and $1000$ for Constraint \ref{constraint_violation}. To allow the HA algorithm to effectively explore the feasibility region, we set the number of search agents to $70$. The HA algorithm stopped when the average change in the fitness function was less than $10^{-8}$ over $30$ iterations.
%
\begin{figure*}
     \centering
     \subfloat[]{\includegraphics[width=.32\linewidth]{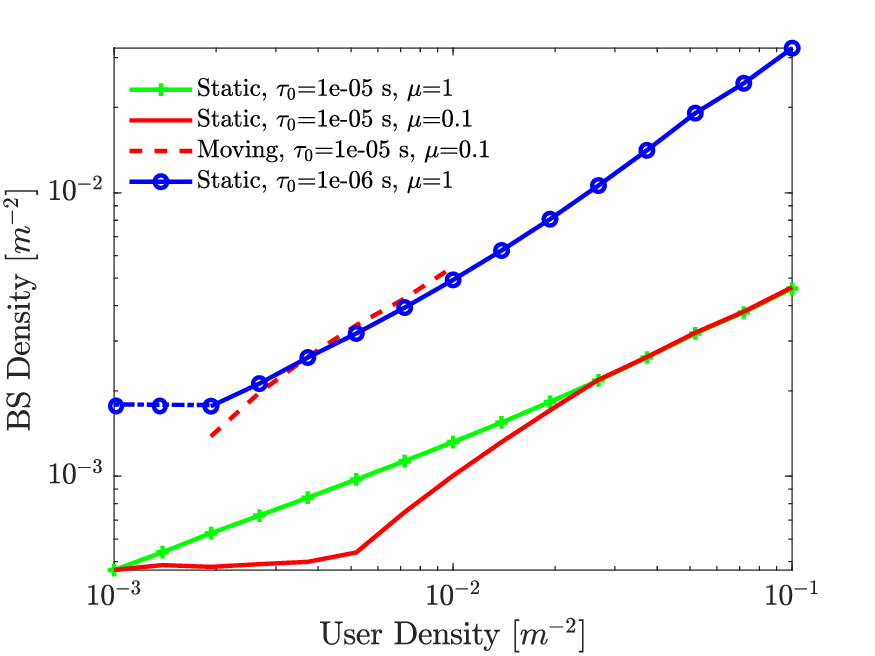}}
     \hspace{-0.028\linewidth}
     \subfloat[]{ \includegraphics[width=.32\linewidth]{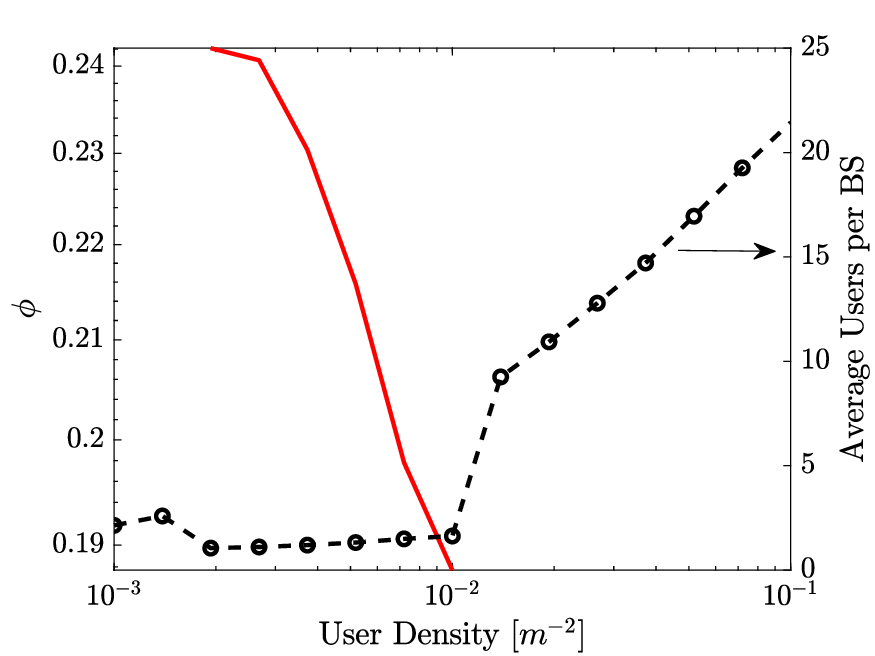}}
    \hspace{-0.01\linewidth}
     \subfloat[]{\includegraphics[width=.32\linewidth]{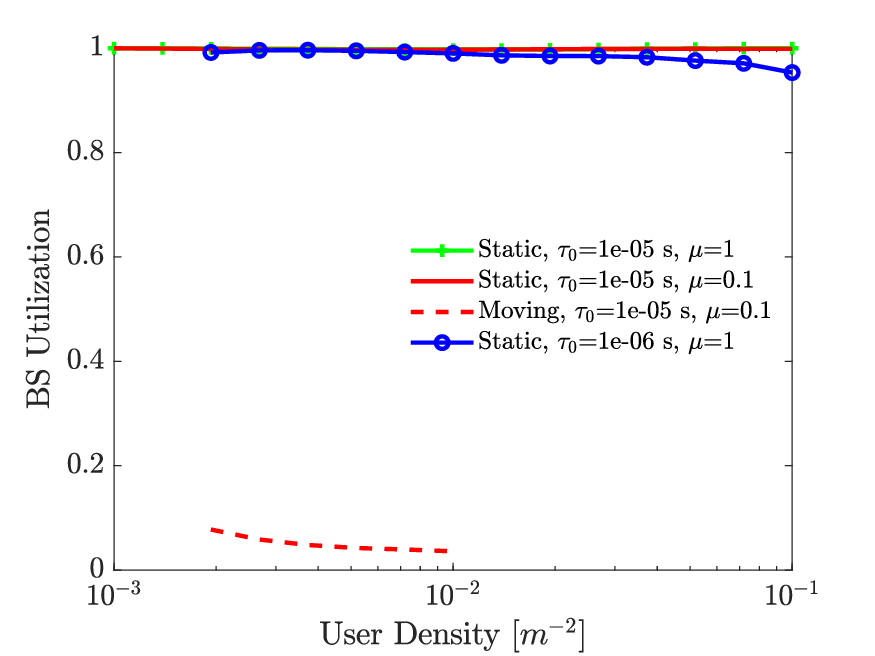}}
     \caption{Network configurations resulting as solutions of the heuristic in Section \ref{mh}, as a function of user density, for different target QoS values, and different relative BS unitary cost $\mu$ (ratio between the unitary cost of MBSs and SBSs). (a) Optimal BS densities; (b) Round robin weight $\phi$ at optimum for backhauling links; (c) Mean BS utilization at optimum.
     }       \label{fig:Results_singleslot}
     \vspace{-.5em}
\end{figure*}
\begin{figure*}
    \centering
     \subfloat[]{\includegraphics[width=.33\linewidth]{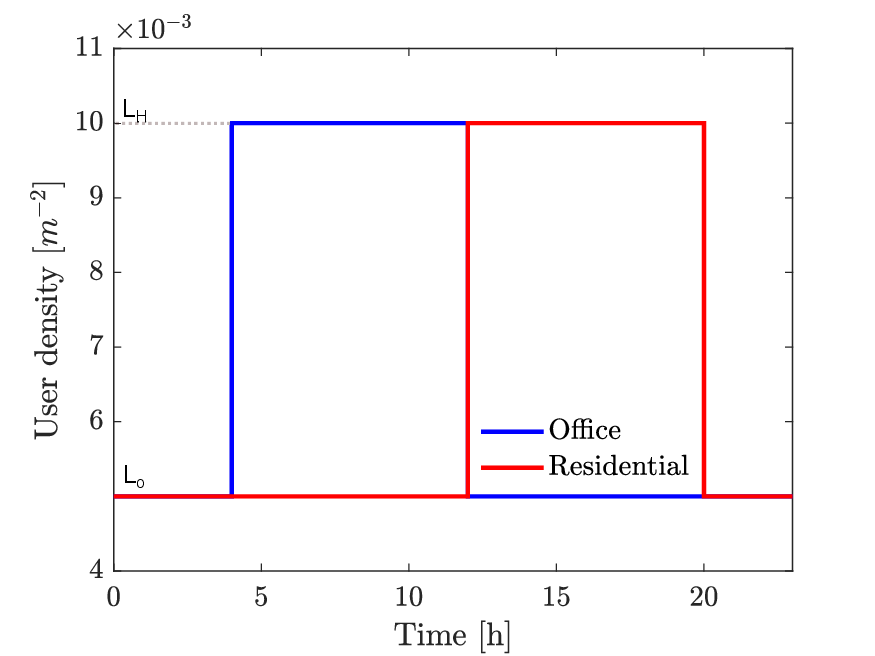}}
     \hspace{-0.028\linewidth}
     \subfloat[]{ \includegraphics[width=.33\linewidth]{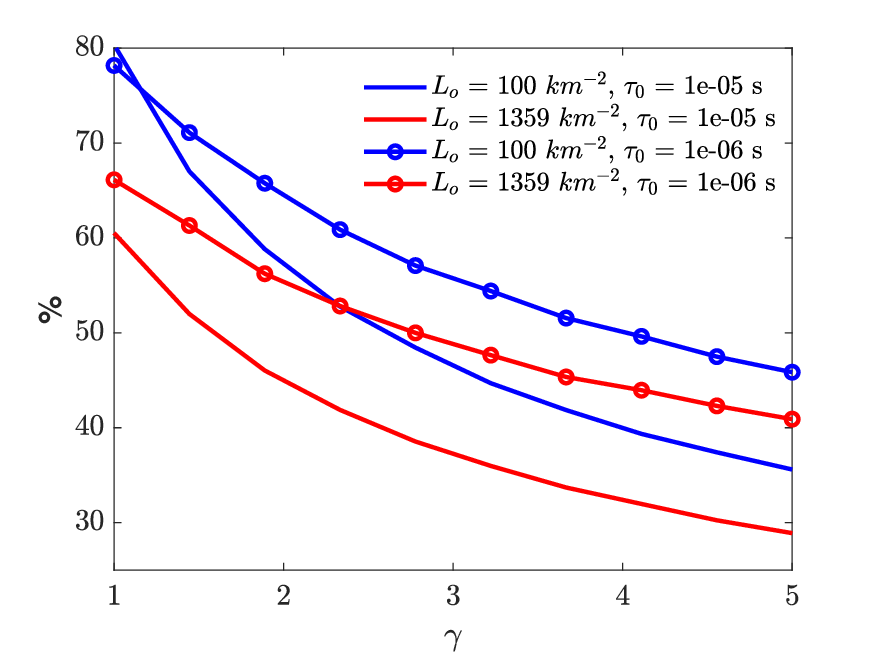}}
    \hspace{-0.028\linewidth}
     \subfloat[]{\includegraphics[width=.33\linewidth]{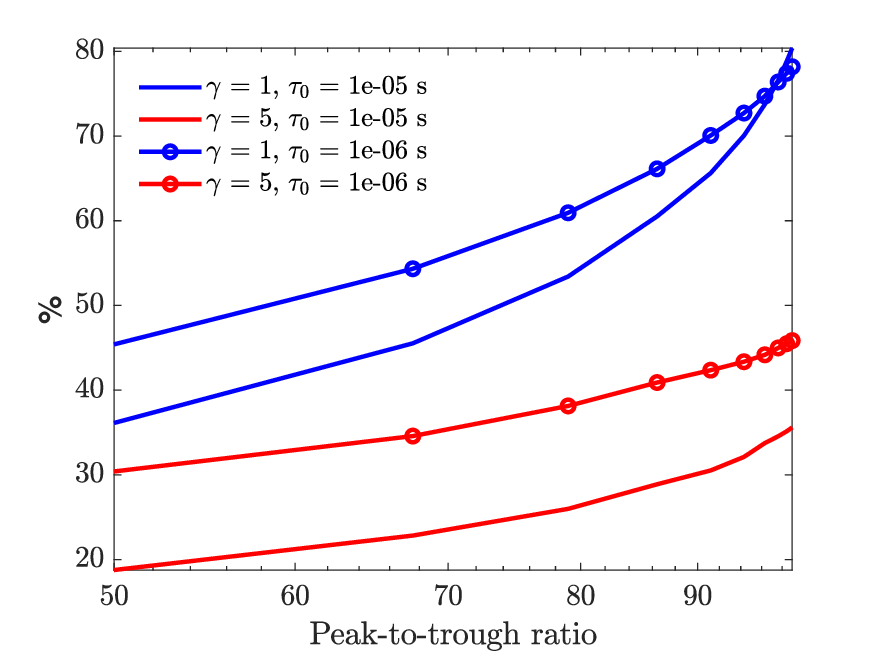}}
     \caption{(a) User density profile over $24$ h for the residential and office district; (b) percentage of BS reused as a function of the area ratio $\gamma$, for a $90\%$ and $95\%$ peak-to-trough ratio, (c) and of peak-to-trough ratio for different values of area ratio.}  
     \label{fig:regions}
     \vspace{-.5em}
\end{figure*}
%
\subsection{Single region static scenario}
To assess the accuracy and reliability of our model, and to identify the main performance patterns arising from it, in a first set of experiments we considered a static scenario, composed of a single region, and we analyzed different setups in terms of user and base station densities while assuming the same transmission power to compare the effects of wireless backhauling on the system.  
As shown in Table \ref{tab_sim}, in all settings the analytical values are within the $95\%$ confidence interval of simulation results, suggesting that our model delivers satisfactory accuracy levels across a variety of settings.  
In addition, the violation probability 
measured from simulations has been always inferior to the target maximum value, set to $5\%$.\\ 
In another set of experiments, 
 we characterized the performance of the systems as resulting from our heuristic as a function of user density, of the target per-bit delay perceived by users. Moreover, as moving and static base stations typically have different installation and deployment ($I\&D$) costs (with wireless backhauling MBS being generally less expensive than SBSs), we evaluated the impact of the relative $I\&D$ 
 unitary cost on the solutions. Our goal has been to understand in which scenarios the typically lower deployment costs of MBSs may compensate for their toll on network resource utilization due to in-band backhauling. \\
As expected, the optimal density of active BSs (Figure \ref{fig:Results_singleslot}(a)) grows with user density in all configurations, at a power law rate which is larger when tightening the constraint on target user-perceived QoS. Indeed, a higher end-user throughput has the double effect of increasing the amount of network resources required both in the fronthaul and in the wireless backhaul of the network. In all settings but one, the optimal configuration is with only SBSs. Only when the relative unitary cost of MBSs is $0.1$ or lower, the optimal solution involves a majority of MBSs for a given range of user densities, with a small density of SBS acting as access points for wireless backhauling links. However, for very low user densities as well as for high ($\geq 10^{-2}$ users per $m^2$) ones, the resource cost of BH links is high enough for the optimal configuration to switch from one in which BB users are mainly served by MBSs, to one with only SBSs. As visible in Figure \ref{fig:Results_singleslot}(b), in this configuration the MBS density and the round-robin weight vary with user density in such a way as to maintain the same (low) mean number of users per BS, to maintain the resource consumption of each BH link at a minimum. The switch to SBS-only happens when the amount of SBS time spent serving BH traffic becomes preponderant (as witnessed by the low value of round-robin weight), so that the toll which MBSs take on the overall network resources makes the configurations with MBSs less competitive than those with only SBSs. 
Overall, these patterns suggest that the resource cost of QoS-aware wireless in-band backhauling has a high impact on the resource efficiency of a MN. 
Note finally that for $\tau_0=10^{-6}$ s for low user densities the optimal BS density does not decrease with decreasing user density, as doing so would not allow satisfying target QoS constraints. Indeed, for such low user densities, delivering the target QoS to users involves keeping a substantial fraction of BSs active, even if at any time instant a significant fraction of them do not serve any user. In those configurations, (proactive) algorithms which activate base stations according to user mobility would be required, but they are out of the scope of this work, and they would not decrease the optimal density of BS to be deployed.\\    
Figure \ref{fig:Results_singleslot}(c) shows the average BS utilizations at the optimum. In those configurations in which only SBS are deployed, utilization is very close to one as this allows minimizing the amount of deployed BSs. When both MBS and SBS are present at the optimum instead, the mean utilization of MBS is substantially lower than that of SBSs. This happens because the mean overall amount of traffic which a SBS has to serve is way larger than that of a MBS, and because, as we have seen, in this configuration the network tries to keep the traffic at each backhauling link at a low level to minimize the amount of resources required to serve it at a SBS. 
The inefficiency of MBS in a scenario where the user distribution is static was expected, since the advantages of MBS can only materialize when their movement allows the same MBS to play the role of multiple SBS in different time intervals. In order to assess the benefit of the multiple roles of MBS we thus need to look at scenarios with more than one region.

\section{Conclusions}
In this paper, we presented a novel analytical framework for quantifying the potential advantages of the MN paradigm within urban environments. 
Our results show that a significant amount of GV as moving base stations can be reused in the majority of the considered setups, suggesting the overall validity of the MN approach to mitigate the need for dense BS deployments.
As a followup, we plan to extend this work to account for more realistic MBS mobility patterns (including the impact of the urban road grid geometry).

\section*{Acknowledgments}
This paper was supported by the UNITY-6G project funded by the European Union’s Horizon Europe Research and Innovation Programme under Grant Agreement N. 101192650, the INTERACT COST Action, project PID2022-140560OB-I00 (DRONAC) funded by MICIU/AEI /10.13039/501100011033 and ERDF, EU and the German Research
Foundation (DFG) within the DyMoNet project under grant
DR 639/25-1.




\newpage
\appendix

\subsection{Proof of \cref{lemma:pdf_distance}}
\label{app:proof_pdf_distance}

    The pdf $f_{m,s}(r)dr$ of the distance of typical user from the serving BS can be obtained as the complementary probability of not having other BS, both of type $s$ and $m$ in the circle of radius $r$ centered at the user. This can be easily computed using the void probability of the concerned PPP. As in general the transmit power of each of the two populations may be different if the distance from a BS of population $s$ is $r$, the distance from a BS of population $m$ such that the received power will be the same as for the BS from population $s$ is given by 
\[
r'=r\rho_{ms}=r\sqrt[\alpha]{\frac{P_s}{P_m}}
\]
The void probability of the PPP associated with the population $s$, evaluated in the ball of radius $r$ centered at the origin is $e^{-\pi r^2\lambda_s}$
Similarly, the probability of not having other BS from the population $m$ at a distance smaller than $r'$ from the user at the origin is given by $e^{-\pi r^2\rho_{ba}^2\lambda_b}$.
Recalling that the two processes are independent, we have:
    \[
    F_{m,s}(r) = 1 - e^{-\pi r^2 (\lambda_s+\lambda_m \rho_{ms}^2)}
    \]
    Thus,
    \[
    f_{m,s}(r) = \frac{d}{dr} F_{m,s}(r) = 2\pi r  (\lambda_s + \lambda_m\rho_{ms}^2) e^{-\pi r^2 (\lambda_s+\lambda_m \rho_{ms}^2)}
    \]

\subsection{Proof of \cref{lemma:interference}}
\label{app:proof_lemma_interference}

\begin{lemma}\label{avg_utilization}
The mean utilization in region $z$ at time slot $t$ of small cells and macro cells is given, respectively, by 
\[
\bar{U}^{j,z}_m=U_m (\bar{\tau}_{m}^{j,z} ) = \frac{\bar{\tau}^{j,z}_{m}}{\tau_0}
\]
and
\[
\bar{U}^{j,z}_s= U_s (\bar{\tau}^{j,z}_{s} ) = \frac{\bar{\tau}_{s}^{j,z}}{\tau_0}
\]
\end{lemma}
\begin{proof} The ideal per-bit delay is the delay perceived by the typical BB user in $x$ when the utilization of his serving small cell is equal to 1. Specifically, 
\[
\tau_{m}^{ideal}(x) = \frac{ N_U + 1}{ C(x)}
\]
where $C(x)$ is the capacity at $x$. When the utilization of the small cell is given by the general expression in \cref{util_MBS}, we can compute the actual per-bit delay of the typical user in $x$ as
\[
\tau^{j,z}_{m}(x) = \frac{ N_U + 1 + \beta_m}{ C(x)}
\]
Thus, the utilization of the small cell serving the typical user in $x$ can be expressed as $U_m(x)\tau^{j,z}_{m}(x)  = \tau_{m}^{ideal}(x)$. The result follows from computing the Palm expectation of both sides and assuming that utilization is tuned such that the average actual per-bit delay is equal to the target value $\tau_0$. The same steps are valid for macro cells. 
\end{proof}
The formula for the average interference is derived by computing the Palm expectation of the total received interfering power $I(D,z,j,k)=\sum_{i \in \phi_j^z}Pr_i^{-\alpha}u_i^z(j)$ at time $t$ and area $z$ for a user distant $D$ from the serving BS, where $P$ is the transmission power depending on the BS tier, $u_i^z(j) =1$ if the BS is active in $z$ at $j$ and $\phi_j^z(k)$ is the PPP thinned by $k$ and intensity $\lambda_{m,j}^z + \lambda_{s,j}^z$.
The final formula is derived from considering that BSs are active only for a fraction of time equal to their utilization and that both static and moving BS contribute to inter-BS interference. Furthermore, we approximate the BS utilization as the average for moving and static BS. Note that this formula also assumes that MBS move by remaining, at any time instant, uniformly distributed on the region of reference.

\subsection{Proof of \cref{th:mean_tau}}
\label{app:proof_th_mean_delay}
We will focus on the derivation of $\bar{\tau}_s^{j,z}$ while the derivation of $\bar{\tau}_m^{j,z}$ follows the same line as in \cite{rizzo2023green}.
The Palm expectation of $\tau_{s}^{j,z}(S(0),D)$ perceived by  user in $S(0)$ (i.e. at the origin) at a distance $D$ from its serving BS, in region $z$ and time slot $j$, is 

\[\E^0[\tau_{s}^{j,z}(S(0),D)] = \E^0\biggl[\frac{N_M(S(0))+\phi_j^z N_{U}(S(0)) }{\phi_j^z C(D(0), P_s,I)} \biggr] \approx \]
\[\int_0^\infty \frac{\E^0[N_M(S(0))+\phi_j^z N_{U}(S(0)) \vert r \leq D \leq r+dr]  }{\phi_j^z C(D(0), P_s,\bar{I}(D))}  \]
\[P(r\leq D\leq r+dr)dr\]
where $\bar{I}(D)$ is the mean interference given by Lemma \ref{lemma:interference}. For $dr \rightarrow 0$ we have:
    $$P(r \leq D \leq r + d r) \approx f_{m,s}(r)$$
with $f_{m,s}(r)$ as in Equation \ref{pdf_distance}.\\ 
To derive the final expression, we compute the expected value of the Poisson distribution of users served by a SBS $E^0[N_{M}+\phi_{j}^z N_{U}  |r \leq D \leq r + d r ]$  of intensity $\phi_{j}^z\lambda_{u,j}^z + \lambda_{m,j}^z$. Thanks to the additivity of the expectation we can split the equation for MBS and BB users. Since the delay for BB users is trivial if only one user is associated with the BS we will consider the truncated PPP. To compute an expression for the average size of the Voronoi cell $h_{BH}(r)$ we proceed as in \cite{rizzo2015}
and we move the typical MBS in $(0,-r)$ so that its serving static BS is located at the origin. Then we consider a user in $(x,\theta)$ served by the BS at the origin and impose that no other BSs are closer. This event occurs with a probability $e^{-\lambda_{s}^zA(r,x,\theta)}$, where $A(r,x,\theta)$ is the area of the circle centered at $(x,\theta)$ that is not overlapped by the circle centered at the typical MBS. The derivation of $h_s(r)$ and $h_m(r)$ is similar but without accounting for the PPP of MBS as serving BS for the BH. Specifically, when the static serving BS is distant $r$, $h_s(r) = \int_0^\infty \int_0^{2\pi} e^{-\lambda_{s}^z A(r,x,\theta) - \lambda_{m,j}^z A(r, x\rho_{ms},\theta)}$ since no other SBS must be closer in $A(r,x,\theta)$ and no other MBS must be in the smaller area $A(r, x\rho_{ms},\theta)$. It is important to note that when computing the average number of BB users and MBS served by the SBS, the sizes of Voronoi cells are different. Specifically, for the BB users we account for MBS in proximity reducing the overall size of the Voronoi cell.  \\
The existence and uniqueness of the fixed point derive from applying the Banach fixed-point theorem to the fixed point problem at hand, as it can be proved that the system of equations for the average per-bit delays is a contraction using well-known inequalities.

\subsection{Proof of \cref{th:cdf}}
\label{app:proof_th_cdf}
For simplicity, we drop the location indication, and we denote $S(0)$ and $D(0)$ as $S$ and $D$ respectively.
$CDF_r(r)$ is the cumulative distribution function of the distance of users from their serving base station derived from the probability of not having other SBS than the serving one in the annulus of radius $r$ and $r+dr$. To prove the result, we 
need to compute the mean per bit-delay in downlink for MBSs as a measurable function $g(r)$. In Euclidean space with Lebesgue measure, measurable functions are continuous. For  $z \in \{R,O\}$ and $j \in \{1,...,J\}$, we rewrite the expression in \cref{eq:user_sbs} for MBS as $$\tau_M = \frac{N_M(S)+\phi_j^z N_{U}(S) }{C(r,P,I)}$$
To obtain $g(r)$, we compute the number of users served by the static BS in $S$ as a function of $r$. For the PPP of users with density $\lambda_{m,j}^z+\phi_j^z \lambda_{u,j}^z$, this quantity has expression 
$(N_M(S)+\phi_j^z N_{U}(S))(r)$. Its mean value is thus given by $(\lambda_{m,j}^z h_{BH}(r)+\phi_j^z \lambda_{u,j}^z h_s(r))$, where $h_s(r)$ and $h_{BH}(r)$ are the mean areas of the Poisson Voronoi cell of a static BS, respectively with and without accounting for MBSs in the system, conditioned to having the user at a distance $r$ from its serving BS and having a SBS as serving BS. Their derivation follows as in the proof of Theorem \ref{th:mean_tau}. 

\subsection{Proof of \cref{thm_violprob}}
\label{app:proof_th_1}
We consider an MBS cell in the region $z$ and interval $j$, serving $N_U$ BB users. We have the following result.
\begin{proposition} For $\lambda_{u,j}^z < 1 $, the variance of the PPP of users is lower than the variance of the Poisson Voronoi cell area.
\end{proposition} 
\begin{proof}
    The variance associated with the number of users in a generic Voronoi cell of area $A$ is equal to the density $\lambda_{u,j}^zA$. Also, if the user density is lower than 1, then the average size of Poisson Voronoi cells associated with BS is $\geq 1\;m^2$. The variance of the surface of Poisson Voronoi cell $A$ can be derived from the second moment 
    as 
    $Var(A) = \frac{2}{7}\Bigl( \frac{1}{\rho_{ms}\lambda_{m,j}^z + \lambda_{s}^z} \Bigr)^2$. Since the ratio between the density of users and BS is always greater than 1, then $(\rho_{ms} \lambda_{m,j}^z + \lambda_{s}^z) < 1$, and the result follows.
\end{proof}
Due to this assumption, the stochasticity of the per-bit delay is only associated with the cell's area rather than being influenced by fluctuations in the number of users.
\begin{lemma} When the utilization of the MBS is less than 1, the PDF  of the per-bit delay of the  backhauling traffic demand in downlink for an MBS is $f_{\tau_d}(t) = a^{\frac{7}{2}}\frac{343}{15}\sqrt{\frac{7}{2\pi}}t^{-\frac{9}{2}}e^{-\frac{7}{2}\frac{a}{t}}$, with $a = \frac{\tau_0}{\lambda_{u,j}^z A}(\rho_{ms}\lambda_{m,j}^z+\lambda_{s}^z)$. 
\end{lemma}
\begin{proof}
We consider the worst case in which users always have data to download. When the utilization of the MBS is less than 1, for the assumption made on the BS service model, the MBS is serving each user with a mean per-bit delay corresponding to the target value, equal to $\tau_0$. Thus the mean per bit delay of the aggregate traffic demand is $\tau_d=\frac{\tau_0}{N_U}$. For Proposition 1, the PDF of $\tau_d$ is dominated by the stochasticity of the area of the cell of the MBS. Thus the random variable $\tau_d$ can be written as  
\begin{equation*}
    \tau_d=\frac{\tau_0}{\lambda_{u,j}^z A}
\end{equation*}
where $A$ is 
the area of the Voronoi cell of the MBSs.
An exact analytical expression of the probability distribution of the area of a Poisson Voronoi tessellation is available only in the single-dimension case. 
However, accurate approximations are available for the two-dimensional case. 
The PDF $f_A(y)$ of the normalized area of a Voronoi cells is
$f_A (y) = \frac{343}{15}\sqrt{\frac{7}{2\pi}}y^{\frac{5}{2}}e^{-\frac{7}{2}y}$, where $y$ is the area of the Poisson Voronoi cell divided by the average surface given by $1/(\rho_{ms}\lambda_{m,j}^z+\lambda_{s}^z)$ where $\rho = \sqrt[\alpha]{{P_m}/{P_s}}$ accounts for the difference in the transmission powers.
The expression for the PDF $f_{\tau_d}$ of the per-bit delay associated with the aggregate traffic demand in downlink is derived from the following transformation formula: 
\[
f_{\tau_d}(\tau) = f_A(g^{-1}(\tau))  \bigg| \frac{d}{d\tau}g^{-1}(\tau) \bigg|
\] 
with $\tau_d = g(A) = \frac{\tau_0}{\lambda_{u,j}^z A}(\rho_{ms}\lambda_{m,j}^z+\lambda_{s}^z)$.
\end{proof}
We now derive the violation probability, i.e. the probability that the per-bit delay perceived by the backhauling connection is insufficient to carry the aggregate traffic demand (that is, that it is larger than $\tau_d$).
Given that with $\tau_M$ we denote the ideal per-bit delay perceived by the backhauling connection, the actual per-bit delay of the backhauling connection is given by
$\tau_M/\bar{U}_s$, where $\bar{U}_s$
   is the mean utilization of SBSs.
We thus have that the violation probability can be expressed as
   
$$ P(\tau_M > \bar{U}_s \tau_d) = 1 - P(\tau_M \leq \bar{U}_s \tau_d) = 1-CDF_{\tau_M}(\bar{U}_s \tau_d)$$
we consider the random variable $ T = \bar{U}_s \tau_d -\tau_M$ and recall that $\tau_d$ and $\tau_M$ are independent from each other. Then we compute the CDF of the new random variable $T$:
$$CDF_T(t) = P(\bar{U}_s \tau_d + (-\tau_M)\leq t) =$$
$$\int_0^\infty \int_{x-t}^{\infty} f_{ \tau_d,\tau_M}(\bar{U}_s x,y)dydx  $$
$$ = \int_0^\infty f_{ \tau_d}(\bar{U}_s x)\int_{x-t}^{\infty} f_{\tau_M}(y)dydx$$
where $f_{\bar{U}_s \tau_d,\tau_M}(\cdot,\cdot)$ is the joint pdf of $(\tau_d,\tau_M)$, the last equality follows from the independence of the given random variables and
$$ \int_{x-t}^{\infty} f_{\tau_M}(y)dy= 1-CDF_{\tau_M}(x-t)$$
Thus, $P(\tau_M > \bar{U}_s \tau_d) = CDF_T(0)$ is given by:
$$\int_0^\infty f_{ \tau_d}(\bar{U}_s x)(1-CDF_{\tau_M}(x-t))dx$$





\end{document}